\documentclass[a4paper]{article}

\begin{document}

\title{Cavity field ensembles from nonselective measurements}
\author{J. Larson and S. Stenholm  \\
Physics Department,\\
Royal Institute of Technology,\\
SCFAB, Roslagstullsbacken 21, S-10691 Stockholm, Sweden\\}
\date{\today}
\maketitle

\begin{abstract}
We continue our investigations of cavity QED with time dependent
parameters. In this paper we discuss the situation where the state
of the atoms leaving the cavity is reduced but the outcome is not
recorded. In this case our knowledge is limited to an ensemble
description of the results only. By applying the Demkov-Kunike
level-crossing model, we show that even in this case, the
filtering action of the interaction allows us to prepare a
preassigned Fock state with good accuracy. The possibilities and
limitations of the method are discussed and some relations to
earlier work are presented.
\end{abstract}

\section{Introduction}
\label{intro}

Cavity QED has become a standard tool of Quantum Optics. It allows
the experimentalists to control both atomic and field variables to
a high degree, and consequently the system can be utilized to
explore many basic features of matter-field interactions. In the
limit when the radiation-induced time scales are fast enough, we
can neglect the various relaxation rates affecting the system, and
then it can be used as a laboratory realization of the well known
Jaynes-Cummings model.

In cavity QED physics, the model has usually been introduced with
time independent parameters. Its utility is, however, preserved if
we allow them to vary at a suitable rate. They should be fast
compared with the relaxation processes of the system but slow
enough not to destroy the mode structure of the cavity. In our
works \cite{jonas1} and \cite{jonas2}, we have considered the new
possibilities offered by such a model.

Because the Jaynes-Cummings model separates into decoupled
two-level systems, we advocated in reference \cite{jonas1} that it
may be used to design a filter mechanism to shape the photon
distribution in the cavity into a desired form. There are earlier
investigations into the effect on the cavity field caused by
recording of the atomic state, see for example \cite{bmg} and
\cite{harel}, but we explore the possibilities deriving from time
dependence of the coefficients. In reference \cite{jonas2}, we
consider the possibility that the desired effects may be derived
simply from the shape of the eigenmodes of the cavity field.

In our earlier works, we have assumed that we record the state of every atom
exiting the cavity after an interaction. In practical situations, however,
the environment rapidly destroys the coherence between the atomic states
after leaving the cavity, even when no measurement is performed. In this case
the state of the field is reduced to one of its possible configurations
without us knowing which one has occurred. We have a situation of
nonselective measurements. After the passage of $m$ atoms through the cavity,
our uncertainty about the actual state of the cavity becomes large. This,
however, is just the situation to be described by a statistical ensembles of
fields, each member occurring with a probability determined from the
sequence of atomic reduction events which brought it about. In this paper we
consider the evolution of the field in such a situation.

The approach may be useful even if we, for some reason, do not
want to record the results of the individual interaction sequence.
In particular, we address the question of creating a pure Fock
state of the field in the cavity. This problem has been discussed
before \cite{walther0}, but all methods tend to become difficult
if one aims at large quantum numbers. Our results may be
considered to be a contribution to the discussion aiming at the
creation of Fock states.

In section \ref{jcmodel}, we set up the problem and introduce the
basic concepts. In section \ref{DKmod} we specialize the
Hamiltonian to the solvable Demkov-Kunike model and discuss its
properties in its adiabatic and non-adiabatic limits. In the
former case we show that it is, ideally, possible to reach
arbitrary $n$-states of the field. Special cases of the latter
situation have been discussed earlier \cite{harel} and
\cite{walther1}, and hence we only make a numerical investigation
of the properties of the result specifically addressing the
question of obtaining a Fock-state. Here the concept of trapping
states \cite{trapp} becomes central, and we suggest a way to
improve the method by velocity selection of the atoms sent into
the cavity. In section \ref{conc} we conclude the presentation
with a discussion.

\section{Adiabatic Jaynes-Cummings model }\label{jcmodel}
We consider a system described by the Jaynes-Cummings model
\cite{jaynes-Cummings}, which is the simplest formulation of the
interaction between matter and a quantized field. The model
assumes that the time of interaction is much shorter than other
time-scales in the system, such as atomic and cavity field
relaxation times. This means that atomic and field losses can be
neglected during the interaction. The Hamiltonian for the full
system is $(\hbar=1)$
\begin{equation}\label{jc}
H_{JC}= \Omega a^{\dagger}a+\frac{\omega}{2}\sigma_3+g\left(a^{\dagger}\sigma^-+a\sigma^+\right),
\end{equation}
where the $\sigma$:s are the ordinary Pauli matrixes and $\{
a,a^{\dagger}\}$ are the Boson operators of the cavity mode.
Within this rotating wave approximation, the number of excitations
is conserved; the operator $N=a^{\dagger}a+\frac{1}{2}\sigma_{3}$
is a constant of motion. We therefore define the interaction
Hamiltonian $H_I=H_{JC}-\Omega N$, which separates the dynamics
into two-level manifolds within the bare states.

The state of the whole atom-field system can be expressed as
\begin{equation}\label{init}
|\Psi\rangle=c_0a_-(0)|0,-\rangle+\sum_{n=1}^{\infty}c_n\Big[a_+(n)|n-1,+\rangle+a_-(n)|n,-\rangle\Big]
\end{equation}
and after the passage of the atom, the state of the system is determined from $a_{\pm}^{\infty}(n)$.
The coefficients $a_{\pm}(n)$ are given by the Schr\"odinger equation
\begin{equation}
i\frac d{dt}\left[
\begin{array}{l}
a_{+}(n) \\
\\
a_{-}(n)
\end{array}
\right] =\left[
\begin{array}{lll}
\displaystyle{\frac{\Delta \omega }{2}} &  & g\sqrt{n} \\
&  &  \\
g\sqrt{n} &  & \displaystyle{-\frac{\Delta \omega }{2}}
\end{array}
\right] \left[
\begin{array}{l}
a_{+}(n) \\
\\
a_{-}(n)
\end{array}
\right] ,  \label{two-ham}
\end{equation}
Here the detuning $\Delta\omega=\omega-\Omega$ and the coupling
$g$ are allowed to depend on time. Initially, the atom is taken to
be either in its upper or lower level
\begin{equation}
\label{initial}
\begin{array}{lllll}
\mathrm{Case\,\, (a):} & & |a_+^0|=1, & & a_-^0=0
\\ & & & & \\
\mathrm{Case\,\, (b):} & & a_+^0=0, & & |a_-^0|=1
\end{array}
\end{equation}
and the initial field distribution is determined from $c_n$.

After the interaction, the atomic state is measured, and from
equation (\ref{init}) it follows that the initial state of the
field is modified by $a_+^{\infty}(n)$ or $a_-^{\infty}(n)$,
depending on whether the atom is found in its upper $|+\rangle$ or
lower $|-\rangle$ state. After $m$ atoms have passed through the
cavity and been measured, the field is multiplied by a series of
$a_{\pm}^{\infty}(n)$:s. It is important to note, that the number
of atoms detected in their upper and lower levels is not enough to
tell what the state of the field is. One also needs the order of
upper and lower detection events. This follows from the fact that,
if the initial and final atomic states are different, then not
only will the field be modified by $a_{\pm}^{\infty}(n)$, but the
photon distribution will also be shifted by one unit. For example,
if an initial lower level atom is found in its upper level after
the interaction, the distribution of the field is
$P_{1,+}(n)=N|a_+^{\infty}(n+1)|^2P_0(n+1)$, where $P_0(n)$ is the
initial photon distribution and $N$ a normalization constant.

Let $\mathbf{k}$ be a vector denoting the sequence of  measured
atoms defined by: $k_j=\pm 1$ if the $j$:th atom's initial and
final measured states have been flipped according to
$|\mp\rangle\,\rightarrow\,|\pm\rangle$, and $k_j=0$ if they are
the same. Then $\mathbf{k}$ uniquely determines the final photon
distribution, provided that the initial state of the field is
known.

If we consider Case (b) in equation (\ref{initial}) and have, for
example, $m=3$ and $\mathbf{k}=(0,1,1)$, that is the first atom is
found in $|-\rangle$ while the other two atoms are found in
$|+\rangle$, then the final photon distribution will have the
shape
\begin{equation}
P_{m,\mathbf{k}}(n)=|a_+^{\infty}(n+1)|^2|a_+^{\infty}(n+2)|^2|a_-^{\infty}(n+2)|^2P_0(n+2).
\end{equation}

If $\nu=\sum_j k_j$, then the photon distribution after $m$ atoms
can be written, in a compact form,  as
\begin{equation}\label{dist}
P_{m,\mathbf{k}}(n)=\frac{1}{N_{\mathbf{k}}}A_{m,\mathbf{k}}(n)P_0(n+\nu),
\end{equation}
where $A_{m,\mathbf{k}}(n)$ is the appropriate sequence of "filter
functions" $|a_{\pm}^{\infty}(n)|^2$ and $N_{\mathbf{k}}$ is the
normalization constant. If the field was initially in a Fock
state, then, whatever the filter functions are, the final field
will also be in a Fock state, shifted by an amount $-\nu$.

If the initial conditions are as Case (a), the probability for
measuring the atomic states $|\pm\rangle$, given a normalized
photon distribution $P=P(n)$, is
\begin{equation}\label{condpro}
\begin{array}{l}
P(+|P)=\sum_n|a_{+}^{\infty}(n+1)|^2P(n)
\\   \\
P(-|P)=1-P(+|P).
\end{array}
\end{equation}
By repeatedly applying the equation above, we generate the
probability $P(\mathbf{k}|P_0)$ of detecting the sequence
$\mathbf{k}$, if the initial photon distribution is $P_0(n)$. Note
that this probability equals the normalization constant in
equation (\ref{dist})
\begin{equation}\label{pronorm}
P(\mathbf{k}|P_0)=N_{\mathbf{k}}.
\end{equation}

Now suppose that we perform a \textit{nonselective measurement} of
the atomic state, i.e. that is we do not record the final state of
the atom. Then the field is described by a density operator,
obtained by tracing over the atomic degrees of freedom
\begin{equation}
\rho^{field}=\mathrm{Tr}_{atom}(\rho)=\sum_{n,n'} c_n
c_{n'}^*\Big[a_+^{\infty}(n)a_+^{\infty*}(n')|n-1\rangle\langle
n'-1|+a_-^{\infty}(n)a_-^{\infty*}(n')|n\rangle\langle n'|\Big] .
\end{equation}
Note that, if we only consider the diagonal elements,
$\tilde{P}(n)=\rho_{nn}^{field}$, we do not need the phases of the
$a_{\pm}^{\infty}(n)$:s. This $\tilde{P}(n)$ is the ensemble
photon distribution as will be explained below. For Case (a) in
equation (\ref{initial}), the initial photon distribution is
$P_0(n)=|c_{n+1}|^2$ and, after $m$ nonselective measurements, the
ensemble photon distribution is obtained from the recurrence
relation
\begin{equation}\label{avfot}
\tilde{P}_m(n)=|a_+^{\infty}(n+1)|^2\tilde{P}_{m-1}(n)+|a_-^{\infty}(n)|^2\tilde{P}_{m-1}(n-1).
\end{equation}
A special case of this equation has been studied before, see
\cite{harel} and \cite{walther1}. It is of essential importance to
point out, that in general $\tilde{P}_m(n)\neq
P_{m,\mathbf{k}}(n)$, for an arbitrary $\mathbf{k}$. The
distribution given in (\ref{avfot}) is the ensemble average over
all possible outcomes $P_{m,\mathbf{k}}(n)$
\begin{equation}\label{avfot2}
\tilde{P}_m(n)=\sum_{\mathrm{All}\,\mathbf{k}}P(\mathbf{k}|P_0(n))P_{m,\mathbf{k}}(n)=\sum_{\mathrm{All}\,\mathbf{k}}A_{m,\mathbf{k}}(n)P_0(n+\nu)=\overline{P_{m,\mathbf{k}}(n)},
\end{equation}
where we have used equations (\ref{dist}) and (\ref{pronorm}) in
the second step and $\overline{P_{m,\mathbf{k}}(n)}$ defines the
ensemble average over all possible $\mathbf{k}$:s. If we consider
cases with a fixed series of outcomes defined by a given
$\mathbf{k}$, we have
\begin{equation}\label{ident}
\begin{array}{lll}
P_{m,\mathbf{k}'}(n)=P_{m,\mathbf{k}}(n),\,\,\,\forall\,\mathbf{k},\,\mathbf{k}' & \Rightarrow &
\tilde{P}_m(n)=P_{m,\mathbf{k}}(n).
\end{array}
\end{equation}
This happens if for example $P(\mathbf{k}|P_0)=1$ for any
$\mathbf{k}$; then this outcome has unit probability and is the
only possible one, all other outcomes have zero probability. The
only possibility that $\tilde{P}_m(n)$ will describe a Fock state
exactly, is if all $P_{m,\mathbf{k}}(n)$ represent the same Fock
state. In other words, if $\tilde{P}(n)$ describes a Fock state,
this agrees with the actual state of the field.

With the above formalism, expectation values $\tilde{\langle
...\rangle}_m$ with respect to the distribution $\tilde{P}_m(n)$
are related to expectation values $\langle
...\rangle_{m,\mathbf{k}}$ for the distribution
$P_{m,\mathbf{k}}(n)$ according to
\begin{equation}
\tilde{\langle ...\rangle}_m=\overline{\langle ...\rangle_{m,\mathbf{k}}}.
\end{equation}
For example, the variance $\tilde{\Delta n}_m^2$ of the distribution $\tilde{P}_m(n)$ is given by
\begin{equation}
 \tilde{\Delta n}_m^2=\overline{\langle n^2\rangle_{m,\mathbf{k}}}-\left(\overline{\langle n\rangle_{m,\mathbf{k}}}\right)^2.
\end{equation}

\section{The Demkov-Kunike model}\label{DKmod}
The Hamiltonian in equation (\ref{two-ham}) is in the most general
form of a $2\times 2$ hermitian matrix \cite{berry}, and has been
widely investigated. For a review of analytically solvable
time-dependent two-level systems see \cite{time}. Here we only
consider the Demkov-Kunike model, in which the two parameters of
the Hamiltonian are given by
\begin{equation}
\begin{array}{l}
\frac{\Delta\omega(t)}{2}=\bar{E}+E_0\tanh\left(\frac{t}{T}\right) \\ \\
g(t)=g_0\,\mathrm{sech}\left(\frac{t}{T}\right).
\end{array}
\end{equation}
If we consider Case (a) of equation (\ref{initial}) and integrate
over time $(-\infty,+\infty)$, the filter functions become
\begin{equation}\label{ffunc}
\begin{array}{l}
|a_+^{\infty}(n)|^2=\left\{
\begin{array}{ll}
1-\frac{\sinh\,\pi T\left(E_0+\sqrt{E_0^2-g_0^2n}\right)\sinh\,\pi T\left(E_0-\sqrt{E_0^2-g_0^2n}\right)}{\cosh\,\pi T\left(\bar{E}+E_0\right)\cosh\,\pi T\left(\bar{E}-E_0\right)} & \bar{E}>E_0
\\ & \\
\frac{\cosh\,\pi T\left(\bar{E}+\sqrt{E_0^2-g_0^2n}\right)\cosh\,\pi T\left(\bar{E}-\sqrt{E_0^2-g_0^2n}\right)}{\cosh\,\pi T\left(\bar{E}+E_0\right)\cosh\,\pi T\left(\bar{E}-E_0\right)} & \bar{E}<E_0
\end{array}\right.
\\ \\
|a_-^{\infty}(n)|^2=1-|a_+^{\infty}(n)|^2.
\end{array}
\end{equation}

These expressions contain the three dimensionless parameters
\begin{equation}\label{para}
\begin{array}{llll}
\Lambda_1=\bar{E}T, & \Lambda_2=E_0T & \mathrm{and} & \eta=g_0T.
\end{array}
\end{equation}
Normally the adiabatic limit arises from large $T$, or
equivalently that, at least one of $\bar{E}$ or $E_0$ becomes
large. In the non-adiabatic limit $\bar{E}\approx E_0\approx 0$.
In the following we consider these two regimes separately.

\subsection{The adiabatic limit}\label{adiabatic}
With the parameters introduced in (\ref{para}) the filter functions may be written as
\begin{equation}\label{filtf}
|a_+^{\infty}(n)|^2=\left\{\begin{array}{ll} 1-\frac{\cosh\pi2\Lambda_2}{2\cosh\pi(\Lambda_1+\Lambda_2)\cosh\pi(\Lambda_1-\Lambda_2)}+\frac{\cos\left(\pi2\sqrt{\eta^2n-\Lambda_2^2}\right)}{2\cosh\pi(\Lambda_1+\Lambda_2)\cosh\pi(\Lambda_1-\Lambda_2)} & \Lambda_1>\Lambda_2 \\   \\ \frac{\cosh\pi2\Lambda_1}{2\cosh\pi(\Lambda_1+\Lambda_2)\cosh\pi(\Lambda_1-\Lambda_2)}+\frac{\cos\left(\pi2\sqrt{\eta^2n-\Lambda_2^2}\right)}{2\cosh\pi(\Lambda_1+\Lambda_2)\cosh\pi(\Lambda_1-\Lambda_2)} & \Lambda_1<\Lambda_2
\end{array}\right..
\end{equation}
For no photons, $n=0$, we have $|a_+^{\infty}(0)|^2=1$. For
\begin{equation}
n>\left(\frac{\Lambda_2}{\eta}\right)^2,
\end{equation}
the last term is an oscillating cosine-term in $n$. In the
adiabatic regime, either $\Lambda_1$ or $\Lambda_2$ becomes large,
and since the denominator grows exponentially in these parameters,
the oscillating cosine-term can be neglected. This means that, for
photon numbers $n>(\Lambda_2/\eta)^2$, the filter functions will
become constant, determined by the first terms/term in
(\ref{filtf}). Then, if the parameters are chosen such that
$\Lambda_2/\eta\approx1$ and at least one of $\Lambda_1$ or
$\Lambda_2$ is large, we have to a good approximation
\begin{equation}\label{adfilt}
|a_+^{\infty}(n)|^2=\left\{
\begin{array}{ll}
1; & n=0
\\ & \\
\kappa ; & n\neq 0
\end{array}\right.,
\end{equation}
where $0\leq \kappa\leq 1$. The number $\kappa$ depends on the
parameters $\Lambda_1$ and $\Lambda_2$. For example, if
$\Lambda_1=\pm\Lambda_2$ then $\kappa=0.5$ or if
$\Lambda_1<\Lambda_2$ (the levels cross) we have $\kappa=0$, while
$\Lambda_2<\Lambda_1$ (no level crossing) gives $\kappa=1$. If
there are non-adiabatic contributions, it is possible to have a
$\kappa$ that differs from the three cases given above. An
important point is that, as long as the process is adiabatic, the
$\kappa=0,\,1$ cases are not sensitive to the parameters of the
problem. On the other hand, the $\kappa=0.5$ case is sensitive to
the condition $|\Lambda_1|=|\Lambda_2|$. The filter function
(\ref{adfilt}) with $\kappa=0.5$ can also be achieved by using a
constant detuning $\Delta\omega$ and an asymmetric field coupling
$g(t)$ in the Hamiltonian (\ref{two-ham}), see \cite{asym}.

With the result (\ref{adfilt}), the equation (\ref{avfot}) becomes
\begin{equation}\label{adfot}
\tilde{P}_m(n)=\kappa\tilde{P}_{m-1}(n)+(1-\kappa)\tilde{P}_{m-1}(n-1).
\end{equation}
This equation can be solved by introducing a generating function
\begin{equation}
G_m(z)=\sum_{n=0}^{\infty}z^n\tilde{P}_m(n).
\end{equation}
The equation (\ref{adfot}) then becomes
\begin{equation}
G_m(z)=\big[\kappa+(1-\kappa)z\big]G_{m-1}(z),
\end{equation}
having the solution
\begin{equation}
G_m(z)=\big[\kappa+(1-\kappa)z\big]^mG_0(z).
\end{equation}
With the field initially in a vacuum, $P_0(n)=\delta_{n,0}$, we find after $m$ atoms the state
\begin{equation}
G_m(z)=\big[\kappa+(1-\kappa)z\big]^m=\sum_{n=0}^{\infty}\left(\begin{array}{c}
m \\ n\end{array}\right)\kappa^{m-n}(1-\kappa)^nz^n.
\end{equation}
As the coefficient of $z^n$ is the photon probability distribution, we find
\begin{equation}\label{binom}
\tilde{P}_m(n)=\left(
\begin{array}{c}
m \\ n
\end{array}\right)
\kappa^{m-n}(1-\kappa)^n.
\end{equation}
$\tilde{P}_m(n)$ is the form of a Binomial distribution with a variance
\begin{equation}
\tilde{\Delta n}_m^2=m\kappa(1-\kappa).
\end{equation}

If $\kappa=0$ we note that we have $\tilde{P}_m(n)=\delta_{n,m}$.
From the discussion above, we know that if the ensemble state of
the field is in a Fock state then the real state of the cavity
field is in the same Fock state. In other words, after $m$ atoms
has passed through the cavity, the state of the field, initially
in vacuum, will be $|m\rangle$. A similar idea for producing a
Fock state has been presented in \cite{haroche2}.

Physically, the process can be understood from the fact that as
each atom traverses the cavity, it is tuned across the field-mode
frequency in an adiabatic manner such that it exits the cavity in
its lower state. Then, after the passage, the mode will contain
one extra photon. On the other hand, if we consider the initial
condition according to Case (b) in equation (\ref{initial})
instead, the mode will be left with one photon less. This holds
for all initial distributions of the field
\begin{equation}\label{erasure}
\begin{array}{lll}
P_m(n)=P_0(n-m) & & \mathrm{for\,\, Case\,\, (a)}
\\ & & \\
P_m(n)=P_0(n+m) & & \mathrm{for\,\, Case\,\, (b)}.
\end{array}
\end{equation}
Obviously, Case (a) will heat up the cavity mode, while Case (b)
cools it down. If lower level atoms are sent through the cavity,
the field mode will eventually end up in the vacuum. The process
can then be used as a "field eraser" \cite{erase}.  Note that the
result in equation (\ref{erasure}) agrees with the one in
(\ref{dist}).

\subsection{The non-adiabatic limit}
If the atomic and field-mode frequencies are on resonance ($\bar{E}=E_0=0$), the filter functions are given by
\begin{equation}\label{resonance}
\begin{array}{c}
|a_+^{\infty}(n)|^2=\cos^2\left(\pi\eta\sqrt{n}\right)  \\ \\
|a_-^{\infty}(n)|^2=\sin^2\left(\pi\eta\sqrt{n}\right),
\end{array}
\end{equation}
where the initial condition has been chosen according to Case (a).
The ensemble distribution (\ref{avfot}), with the filter functions
as in equation (\ref{resonance}), has been investigated in
\cite{harel} and \cite{walther1}, so we will not study it in great
detail here. However, we specifically investigate its prospects to
speed up the production of a Fock state.
%
%
Figure \ref{fig1} shows how the distribution $\tilde{P}_m(n)$,
initially in a coherent state with $\bar{n}=47$, evolve as the
number of atoms $m$ increases. The filter functions are given by
(\ref{resonance}) with $\eta=1$. We see how the distribution
builds up around photon numbers $n'=35,\,48$ and $63$. These
$n'$:s fulfill the trapping condition \cite{trapp}
\begin{equation}\label{trap}
\begin{array}{ll}
\sqrt{(n'+1)}\eta=q, & q=1,\,2,\,3,\,...\,.
\end{array}
\end{equation}
The corresponding states $\{|n'\rangle\}$, fulfilling the trapping
condition (\ref{trap}), are called  trapping states and are
labelled by $(n',q)$. If the field is in a trapping state, then,
since $|a_+^{\infty}(n'+1)|^2=1$,  all atoms will exit the cavity
in their upper level and the state of the field is just multiplied
by a phase-factor. The integer $q$ is the number of the
corresponding maximum of  $|a_{+}^{\infty}(n')|^2$, in other words
the index of the trapping state. For a given trapping state
$(n',q)$, there are $q-1$ other trapping states with photon
numbers smaller than $n'$. $q$ also gives the number of Rabi
cycles during the interaction \cite{walther2}. When the atomic
velocity can be controlled and consequently the effective
interaction time $T\propto\eta$, we may change the number of Rabi
cycles $q$ in the trapping state $(n',q)$, since $T\sim q$.

Since $|a_-^{\infty}(n'+1)|^2=0$ for the trapping states, it is
clear from (\ref{avfot}) that the states $|n'\rangle$ are not
coupled to the states $|n'+1\rangle$. This results in a separation
of the space $\{|n\rangle\}$ into disconnected blocks; population
within one block will not leak into another. If $q$ is small, the
distance between two consecutive trapping states is large, and may
exceed the width of the initial photon distribution. Then, if this
distribution lies in one such block between two maxima of
$|a_{+}^{\infty}(n')|^2$, it will eventually end up in the
trapping state with the largest photon number $n'$. In this way,
the situation allows us to create any Fock state. However, using
this method, the preparation of Fock states with large photon
numbers is difficult for two reasons:

\begin{tabular}{l p{10 cm}}
\\
1. &  
When the photon distribution "approaches" the trapping state, more and more atoms will exit the cavity in their upper state and leave the field approximately unchanged. This follows from the small curvature of the of the filter functions (\ref{resonance}) around one maximum with a large $n'$ and a small $q$. The larger the photon number $n'$ the smaller the curvature, and the number of atoms needed for creating the Fock state is growing rapidly. This means that relaxation of the field may not be negligible.  \\  & \\

2. & 
The error in the atomic velocity, or in the interaction time $T$, propagates to $n'$. When this error is of the order of unity, the trapping condition is clearly destroyed. This undesired effect manifests itself for large photon numbers, since the uncertainty in $n'$ grows linearly with the photon number.
\end{tabular}
\\

The small curvature of the filter functions slows down the
process, so if the curvature could be increased as the
distribution approaches the trapping state, the number of atoms
needed would be reduced. This can be achieved by changing the
atomic velocity, and consequently the trapping state, in such a
way that $(n',q)\rightarrow(n',q')$, where $q'>q$.

In figure \ref{fig2}, we have plotted $\tilde{P}_m(10)$, when the
initial state was a coherent one with $\bar{n}=4$. The solid line
(a) gives the probability when all atoms have the same velocity,
such that $n'=10$ fulfills the trapping condition (\ref{trap})
with $q=1$. For the solid line (b), the velocities of consecutive
atoms are chosen such that $q$ is increased by one unit for each
atom. Starting with $q=1$ it means that the first atom has a
velocity in which the state $(n',q)=(10,1)$ fulfills the trapping
condition (\ref{trap}), and the velocity of the second atom is
changed such that $(10,2)$ instead satisfies the condition. The
trapping states thus occur according to
\begin{equation}
(10,1)\rightarrow(10,2)\rightarrow(10,3)\rightarrow...
\end{equation}
for each atom. This sequence may not be the most efficient one,
but it clearly shows how the process is speeded up. The other two
lines (dot-dash) in figure \ref{fig2} describe the same processes
as the ones above, but with a 2 $\%$ Gaussian random error in the
effective interaction time $T$. The error in $T$ clearly "washes
out" the trapping condition as mentioned above.
%
%
\section{Conclusion}\label{conc}
In this paper we have continued our investigation into the
possibilities offered by letting the parameters of an atom-cavity
system depend on time. It is essential that this variation is slow
enough to retain the identity of the cavity modes and atomic
states; adiabatic changes are assumed. On the other hand, the
interaction times considered have to be short enough that
relaxation processes can be neglected.

In our earlier publications we have investigated how such time
dependence combined with observation of the state of the outgoing
atoms may be used as a photon state filter. Choosing a suitable
time variation, desired photon states may be achieved. In this
paper we continue the investigation by assuming reduction of the
atomic state but no recording of the result. This may take place
spontaneously or by special design. The result is, however, that
knowledge about the cavity state is assembled into a density
matrix, representing the ensemble of events compatible with the
events having occurred. In general this conveys less information
about the state than a selective measurement does. However, if the
method can be used to prepare a pure Fock state, the result
becomes uniquely defined.

The analysis utilizes a Jaynes-Cummings model with time dependent
coefficients. This splits the problem into blocks that can be
analysed separately. Using the well known level crossing model
presented by Demkov and Kunike, we may look separately at the
adiabatic and non-adiabatic limits of this model. Our numerical
investigations suggests some scaling of the process:

In the adiabatic limit it takes $n$ atoms to prepare the $n:$%
th Fock state. In the non-adiabatic limit and with the trapping
state fixed, the number of atoms grows with $n^2$ which may
conflict with the requirement of no relaxations occurring. If,
however, the consecutive atoms utilize consecutive trapping
states, i.e. their velocities are fine tuned before entering the
cavity, the number of atoms may be reduced to $n$ again.

We have not been able to solve the general problem: given that we want to
reach an arbitrarily chosen final photon state, is there a time dependence
which guarantees this? We have only illustrated the utility of the method
for the preparation of a selected Fock state.

\newpage

\begin{figure}
\caption {The probability distribution $\tilde{P}_m(n)$ is plotted
versus the the number of atoms $m$. Here the dimensionless
parameter $\eta=1$. The distribution is built up around the photon
numbers $n=35,\,\,48$ and $63$. These photon numbers all fulfill
the trapping condition (\ref{trap}). } \label{fig1}
\end{figure}

\begin{figure}
\caption {This figure shows how the probability $\tilde{P}_m(10)$,
i.e. having $n=10$, evolves with the number of atoms $m$. For the
solid line (a), the velocities of the atoms are all the same and
chosen such that $n=10$ fulfills the trapping condition
(\ref{trap}) with $q=1$. For line (b), the velocities are changed
for each atom in such a way, that the trapping states $(n,q)$
occur in the sequence
$(10,1)\rightarrow(10,2)\rightarrow(10,3)...$, for consecutive
atoms. The two dot-dash lines are the same as the solid lines, but
with a 2 $\%$ Gaussian random error in the atomic velocity and
consequently the interaction time. The spread in velocities
clearly destroys the trapping effect. } \label{fig2}
\end{figure}

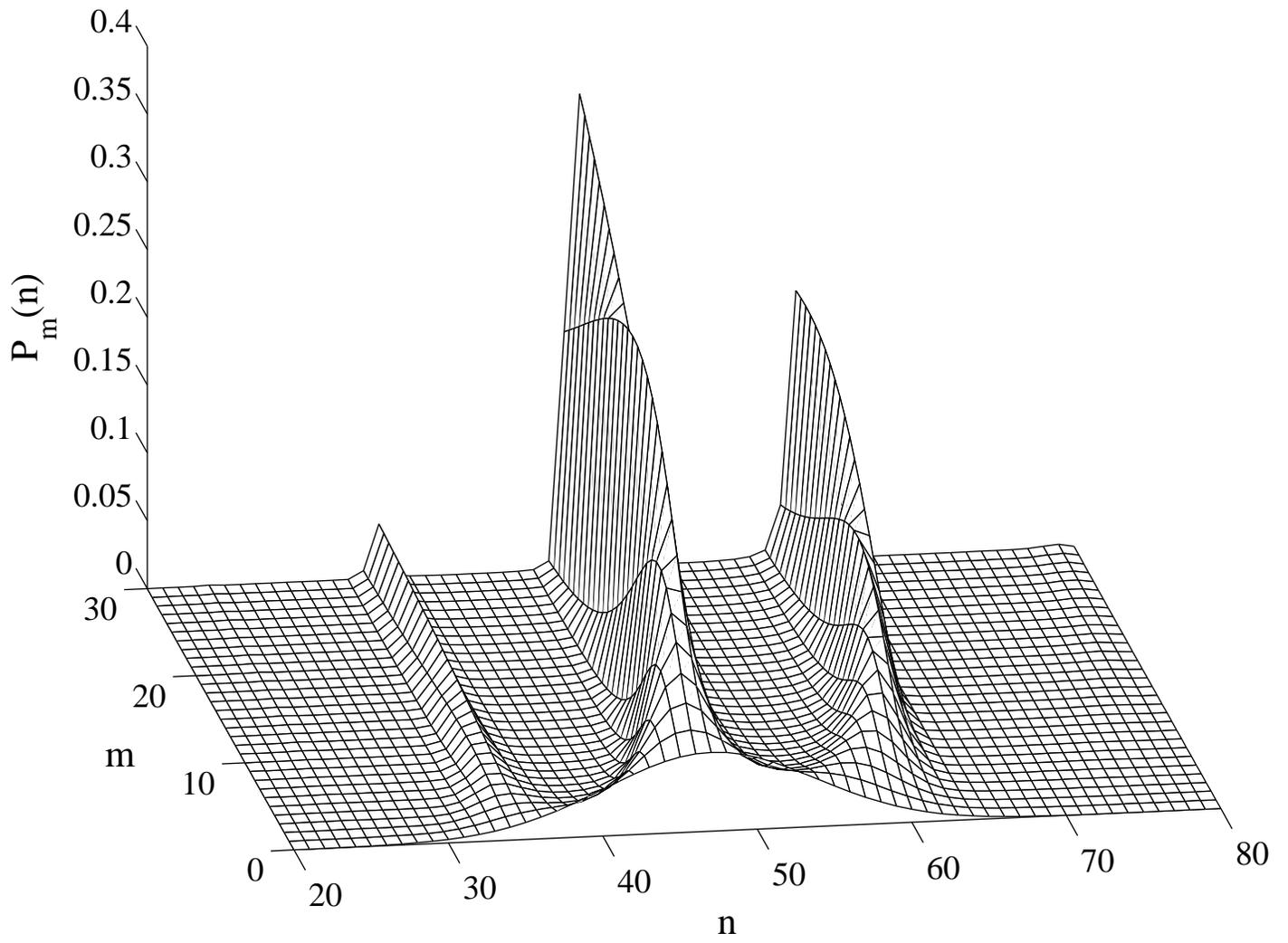
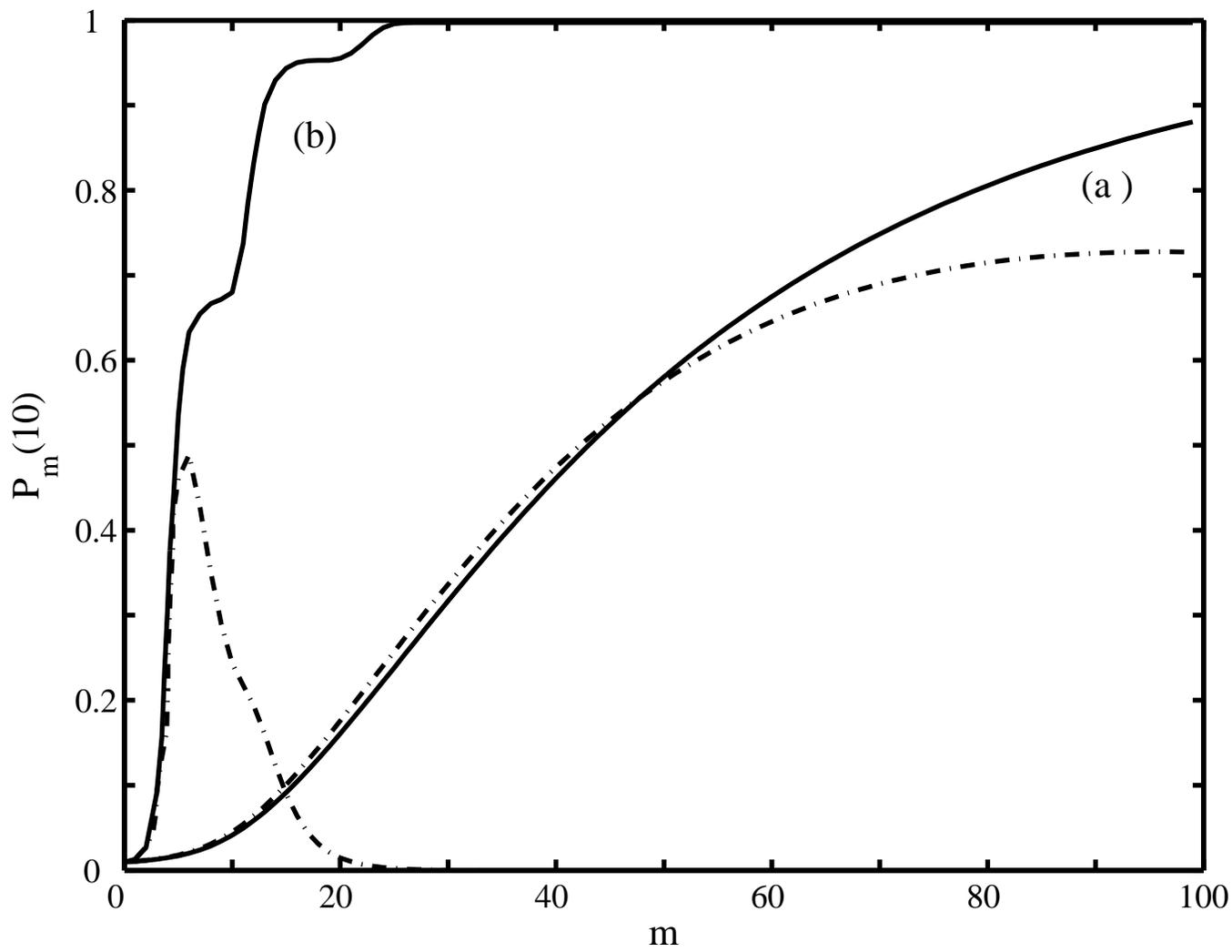
\end{document}